\documentclass[structabstract]{aa}
\usepackage{graphicx}
\usepackage{hyperref}
\hypersetup{
    colorlinks,
    citecolor=blue,
    filecolor=blue,
    linkcolor=blue,
    urlcolor=blue,
    menucolor=black}
\usepackage{txfonts}
\begin{document}

\title{The relationship between bipolar magnetic regions and their sunspots}
\titlerunning{BMRs and their sunspots}
\author{K. L. Yeo\inst{1} \and S.~K.~Solanki\inst{1,2} \and N.~A.~Krivova\inst{1} \and J.~Jiang\inst{3,4}}
\institute{Max-Planck Institut f\"{u}r Sonnensystemforschung, Justus-von-Liebig-Weg 3, 37077 G\"{o}ttingen, Germany\\\email{yeo@mps.mpg.de}
\and{}School of Space Research, Kyung Hee University, Yongin, 446-701 Gyeonggi, Korea
\and{}School of Space and Environment, Beihang University, Beijing 100191, China
\and{}Key Laboratory of Solar Activity, National Astronomical Observatories, Chinese Academy of Sciences, Beijing 100012, China}
\abstract
{The relationship between bipolar magnetic regions (BMRs) and their sunspots is an important property of the solar magnetic field, but it is not well constrained. One consequence is that it is a challenge for surface flux transport models (SFTMs) based on sunspot observations to determine the details of BMR emergence, which they require as input, from such data.}{We aimed to establish the relationship between the amount of magnetic flux in newly emerged BMRs and the area of the enclosed sunspots, and examine the results of its application to an established SFTM.}{Earlier attempts to constrain BMR magnetic flux were hindered by the fact that there is no extensive and reliable record of the magnetic and physical properties of newly emerged BMRs currently available. We made use of the empirical model of the relationship between the disc-integrated facular and network magnetic flux and the total surface coverage by sunspots reported in a recent study. The structure of the model is such that it enabled us to establish, from these disc-integrated quantities, an empirical relationship between the magnetic flux and sunspot area of individual newly emerged BMRs, circumventing the lack of any proper BMR database.}{Applying the constraint on BMR magnetic flux derived here to an established SFTM retained its key features, in particular its ability to replicate various independent datasets and the correlation between the model output polar field at the end of each cycle and the observed strength of the following cycle. The SFTM output indicates that facular and network magnetic flux rises with increasing sunspot magnetic flux at a slowing rate such that it appears to gradually saturate. This is analogous to what earlier studies comparing disc-integrated quantities sensitive to the {amount} of faculae and { network} present to sunspot indices had reported. The activity dependence of the ratio of facular and network flux to sunspot flux is consistent with the findings of  recent studies:  although the Sun is faculae-dominated (such that its brightness is mostly positively correlated with activity), it is only marginally so as facular and network brightening and sunspot darkening appear to be closely balanced.}
{}
\maketitle

\newcommand{\abmr}    {A_{\rm BMR}}
\newcommand{\abmrf}   {A_{\rm BMR,F}}
\newcommand{\abmrs}   {A_{\rm BMR,S}}
\newcommand{\abmrx}   {\tilde{A}_{\rm BMR}}
\newcommand{\atotbmrs}{\Sigma{}\abmrs}
\newcommand{\atots}   {\Sigma{}A_{\rm S}}
\newcommand{\atotfn}  {\Sigma{}A_{\rm FN}}
\newcommand{\bmax}    {B_{\rm max}}
\newcommand{\bmaxx}   {\tilde{B}_{\rm max}}
\newcommand{\fbmr}    {\Phi_{\rm BMR}}
\newcommand{\fbmrf}   {\Phi_{\rm BMR,F}}
\newcommand{\fbmrs}   {\Phi_{\rm BMR,S}}
\newcommand{\ffn}     {\Phi_{\rm FN}}
\newcommand{\fspot}   {\Phi_{\rm S}}
\newcommand{\ftotfn}  {\Sigma\ffn}
\newcommand{\ftots}   {\Sigma\fspot}
\newcommand{\ftotbmrf}{\Sigma\fbmrf}
\newcommand{\latlon}  {\left(\lambda,\phi\right)}
\newcommand{\latlonp} {\left(\lambda_{+},\phi_{+}\right)}
\newcommand{\latlonm} {\left(\lambda_{-},\phi_{-}\right)}
\newcommand{\latlonpm}{\left(\lambda_{\pm},\phi_{\pm}\right)}
\newcommand{\logrhk}  {\log{}R'_{\rm HK}}
\newcommand{\sep}     {\Delta\beta}
\newcommand{\sepsq}   {\left(\sep\right)^2}
\newcommand{\ns}      {\eta_{\rm S}}
\newcommand{\nr}      {\eta_{\rm R}}

\section{Introduction}
\label{introduction}


The global solar dynamo produces the cyclic emergence of bipolar magnetic regions (BMRs) {on the solar surface} which underlies the 11-year activity cycle {\citep{charbonneau20}}. Emergent BMRs range from the larger sunspot-bearing magnetic bipoles, which go on to form active regions, to the smaller spot-free and relatively short-lived (timescale of hours) ephemeral regions (ERs) \citep{harvey93a,harvey00,hagenaar01,hagenaar03,thornton11}. How the larger BMRs relate to the sunspots they enclose and how ER emergence relates to the {{number of}} sunspots present {are of interest as they connect} directly to how solar magnetism and activity relate to sunspot prevalence, with implications for the study of a wide range of solar phenomena. 


The large-scale photospheric magnetic field evolves as BMRs emerge, their magnetic { flux is} transported and dispersed by systematic and turbulent motions, and magnetic elements of opposite polarity approach and cancel each other \citep{vandrielgesztelyi15}. This process can be described via the surface flux transport equation \citep{leighton64,devore84} or, at a simpler level, by a closed system of ordinary differential equations \citep{solanki00,solanki02}. Models have been developed to reproduce the past and forecast the future evolution of the large-scale photospheric magnetic field by solving the transport equation for a given set of initial conditions and details of BMR emergence, aptly termed surface flux transport models (SFTMs) \citep{jiang14}. Some studies have simulated the evolution over the past few centuries by prescribing to SFTMs the details of BMR emergence over this period, {{as}} determined from sunspot number or sunspot area records \citep[e.g.][]{wang05,cameron10,jiang11a,jiang11b}. Sunspot number and sunspot area records are employed as they are the only direct observations of solar surface magnetism to go this far back in time. {Otherwise, SFTMs usually make use of solar magnetograms \citep[e.g.][]{whitbread18,virtanen18,yeates20} or Ca II K spectroheliograms \citep[e.g.][]{virtanen19} instead. While magnetogram and Ca II K spectroheliogram observations do not extend as far back in time as sunspot observations, the details of BMR emergence can be established more directly and accurately from such data.}

Since the large-scale photospheric magnetic field is the observable part of the global dynamo and determines the structure of the coronal magnetic field, SFTMs represent a powerful tool in the study of these solar phenomena \citep{cameron12,mackay12,jiang13}. Models of solar irradiance variability are often employed to generate the historical solar forcing input required by climate simulations \citep{haigh07} and to investigate how the brightness variability of the Sun compares to that of other cool stars \citep{shapiro14,shapiro15,shapiro16,nemec20b,nemec20a}. Through their incorporation into models of solar irradiance variability, discussed next, SFTMs are also relevant to the study of the Earth's climate and of the solar-stellar connection.


{Solar irradiance variability on timescales greater than a day is believed to be mainly driven by} the variation in the prevalence and distribution of dark sunspots and bright faculae and { network} on the solar disc as the photospheric magnetic field evolves \citep{solanki13,yeo17b}. Accordingly, solar irradiance models describe the variation in this quantity due to the intensity deficit from sunspots and the excess from faculae and { network}, and this information is determined from observations of solar magnetism \citep{domingo09,yeo14b}.
 
The variation in solar irradiance over the past few centuries, a period of particular interest to climate studies, is often modelled from sunspot number and sunspot area records  \citep[e.g.][]{lean00,krivova07,krivova10,dasiespuig14,dasiespuig16,coddington16,wu18}. Modelling solar irradiance variability with sunspot observations means inferring not just the effect of sunspots, but also the effect of faculae and { network} on solar irradiance from information about sunspots alone. This can be aided by SFTMs aimed at reproducing the evolution of the large-scale photospheric magnetic field from the same sunspot observations. For example, \cite{coddington16} {incorporated into their solar irradiance reconstruction the decadal variability in the SFTM} reported by \cite{wang05}. In another instance, the solar irradiance reconstruction by \cite{dasiespuig14,dasiespuig16} is based on synthetic full-disc magnetograms generated from {sunspot area and sunspot number records using} the  SFTM of \cite{cameron10} and \cite{jiang11a,jiang11b} (hereafter CJSS).

Various studies found the facular area to increase with increasing sunspot area at a diminishing rate such that it appears to gradually saturate \citep{foukal93,foukal96,foukal98,chapman97,shapiro14}. {\cite{solanki98} and \cite{yeo20} (hereafter  YSK){{,}} compared various disc-integrated quantities that are sensitive to the amount of faculae and { network} present to sunspot area and/or sunspot number and returned analogous results.} As first pointed out by \cite{foukal93}, these results suggest that the balance between the competing effects of bright faculae and { network} and of dark sunspots on solar irradiance might shift towards the latter as we go from low to high activity levels. {This is believed to extend to cool stars in general,} such that as we go from less to more active stars, we transit from a faculae-dominated regime where facular and network brightening dominates starspot darkening to a spot-dominated regime where the converse is true. {This explains} why less active stars appear to become brighter with increasing activity, while more active stars exhibit the opposite trend \citep{lockwood07,hall09,shapiro14,radick18}. The balance between bright faculae and { network} and dark sunspots on the solar disc, and how that might change with the activity level and the inclination of the Sun to the observer, is of interest for the relevance to the understanding of the brightness variability of the Sun and other cool stars. {In recent studies} \cite{nemec20b,nemec20a} extended the model of solar irradiance variability by \cite{dasiespuig14,dasiespuig16}, which incorporates the SFTM by CJSS, {to probe the inclination-dependence of this balance.}


In order to reproduce the evolution of the large-scale photospheric magnetic field over a given period of time, SFTMs require as input the physical properties of each BMR that had emerged over this time interval, including the timing, location, and amount of magnetic flux at emergence. However, deriving {this information} from observations of sunspots  is less than straightforward. {As ERs do not contain or emerge synchronously with sunspots and are difficult to observe due to their relatively small sizes and short lifetimes,} the relationship between ER emergence and sunspot prevalence is not exactly known. As for sunspot-bearing BMRs, it is less than straightforward to infer the details of their emergence from sunspot number records {as these are essentially sunspot and sunspot group counts, and therefore do not contain information on individual sunspot groups.} While sunspot area records do tabulate the location and area of individual sunspot groups, how the amount of magnetic flux in a given newly emerged sunspot-bearing BMR, $\fbmr$, depends on the area of the enclosed sunspots, $\abmrs$, is not well constrained. By the term `newly emerged' we refer to the point where the sunspot-bearing BMR has fully emerged but not yet started to decay. This paper   focuses  on the issue of the relationship between $\fbmr$ and $\abmrs$. For the sake of brevity, hereafter  the term BMR denotes sunspot-bearing BMRs alone, excluding ERs.

The difficulty in establishing BMR emergence from sunspot observations leaves SFTMs to make varying assumptions. The \cite{wang05} model, based on the \cite{hoyt98} sunspot group number record, accommodated the fact that there is no information on individual sunspot groups in such data by assigning the same amount of magnetic flux to each of the emergent BMRs in a given solar cycle. In the CJSS model,  detailed in Sect. \ref{sftm1}, {the $\abmrs$ of each BMR is taken directly from sunspot area records. We let $\sep$ denote the separation between the two poles of a BMR. The model combines an empirical relationship between $\sep$ and $\abmrs$ and the assumption that $\fbmr$ is linearly proportional to $\sepsq$ to determine $\fbmr$ from $\abmrs$.}


Few studies have attempted to constrain {the relationship between $\fbmr$ and $\abmrs$. From an examination of the faculae in active regions and newly emerged BMRs, \cite{schrijver94} postulated that $\fbmr$ scales linearly with the total area extended by the faculae and sunspots in each BMR, $\abmr$. This} can be combined with the empirical relationship between facular area and sunspot area \citep{foukal93,foukal96,foukal98,chapman97,shapiro14} to roughly relate $\fbmr$ to $\abmrs$. In a more recent study, \cite{lemerle15} {compared the database of newly emerged BMRs compiled by \cite{wang89a} to the USAF-NOAA sunspot area record and} found $\fbmr$ to conform to a  power-law relationship with $\abmrs$ with a power-law exponent of 0.71. The analyses of \cite{schrijver94} and \cite{lemerle15} (hereafter  SH and LCC){{,}} {were restricted by certain limitations of the observations available to them, discussed in Sect. \ref{compareshlcc}. It would be straightforward to derive the relationship between $\fbmr$ and $\abmrs$ from an extensive and reliable record of the magnetic and physical properties of newly emerged BMRs. There is, however, no such record currently available, {{as}} also discussed in Sect. \ref{compareshlcc}.}


In this study we aim to establish the relationship between the amount of magnetic flux in each newly emerged BMR, $\fbmr$, and the area of the enclosed sunspots, $\abmrs$. We make use of the empirical model relating the disc-integrated facular and network magnetic flux, $\ftotfn$, to the total surface coverage by sunspots, $\atots$ presented by YSK \citep{yeo20}, which builds on the earlier work by \cite{preminger05,preminger06a,preminger06b,preminger07}{, abbreviated here as PW.} {As we  demonstrate, the set-up of the YSK model allows us to establish from it an empirical relationship between $\fbmr$ and $\abmrs$ while circumventing the lack of a proper database of newly emerged BMRs.} In the following, we  describe the YSK model (Sect. \ref{model1}), derive the empirical relationship between $\fbmr$ and $\abmrs$ (Sect. \ref{model2}), and compare it to the SH and LCC results (Sect. \ref{compareshlcc}). Then, we apply each of these constraints on $\fbmr$ to the SFTM by CJSS (Sect. \ref{sftm1}) and examine the effect on the model output (Sect. \ref{sftm2}). Making use of the surface flux transport simulation based on the empirical relationship between $\fbmr$ and $\abmrs$ derived here, we will look at the balance between the amount of faculae and { network} and the amount of sunspots on the solar disc, and how that changes with solar activity level (Sect. \ref{discussion}). Finally, we provide a summary of the study in Sect. \ref{summary}.

\section{The relationship between the amount of magnetic flux in newly emerged bipolar magnetic regions and the area of the enclosed sunspots}
\label{model}

\subsection{\cite{yeo20} model}
\label{model1}


In this section we derive an empirical relationship between the amount of magnetic flux in each newly emerged bipolar magnetic region or BMR, $\fbmr$, and the area of the enclosed sunspots, $\abmrs$. { For this purpose we employ the empirical model of the relationship between the disc-integrated facular and network magnetic flux, $\ftotfn$, and the total surface coverage by sunspots, $\atots$, recently reported by YSK \citep{yeo20}.} The YSK model is given by
\begin{equation}
\ftotfn=\left(\atots\right)^{h_1}\otimes{}H+h_2,
\label{eqnysk1}
\end{equation}
where $h_1$ and $h_2$ are fit parameters. The convolution of $\left(\atots\right)^{h_1}$ with $H$, a finite impulse response (FIR) filter, describes the variation in $\ftotfn$ associated with sunspot-bearing BMRs. The $h_2$ term represents network magnetic flux that did not originate from sunspot-bearing BMRs;   it is assumed in the model to be constant. The FIR filter is given by
\begin{equation}
H\left(t\right)=h_3\max\left[\exp\left(-\frac{|t|}{h_4}\right)\cos\left(\frac{2\pi{}t}{t_\sun}\right),0\right],
\label{eqnysk2}
\end{equation}
where $-0.25t_\sun\leq{}t\leq14.25t_\sun$, $t_\sun=26.24\ {\rm days}$ denotes the synodic rotation period of the Sun, and $h_3$ and $h_4$ are fit parameters.

\begin{figure}
\centering
\includegraphics{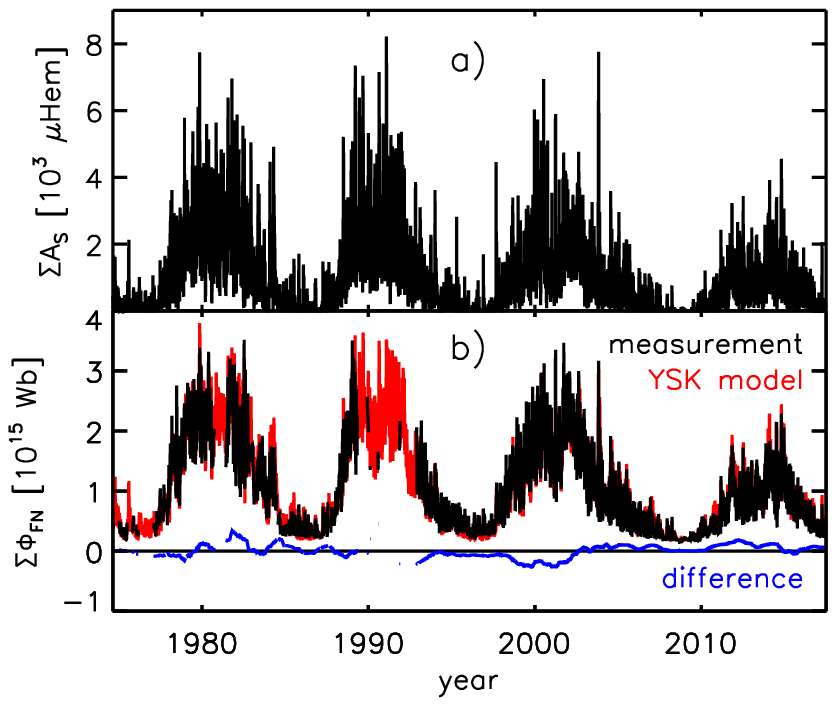}
\includegraphics{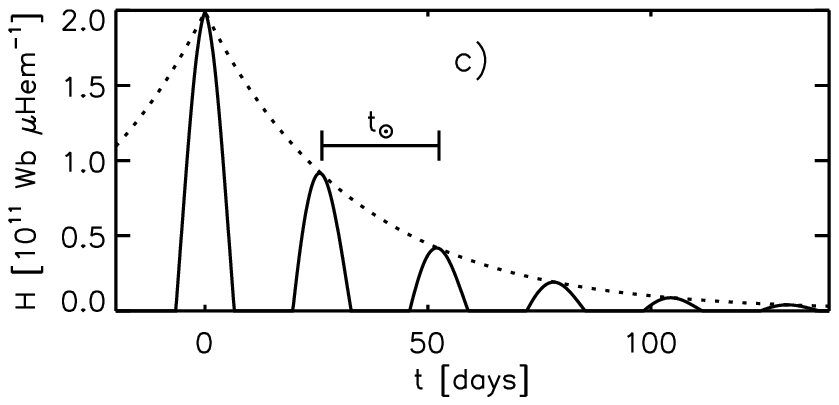}
\caption{{The observed $\atots$ and $\ftotfn$, the YSK model (Equation \ref{eqnysk1}) reconstruction of $\ftotfn$ from $\atots$ and the $H$ term in this model (Equation \ref{eqnysk2}).} a) Composite time series of {{the total surface coverage by sunspots,}} $\atots${{,}} by \cite{balmaceda09}. b) Composite time series of {{the disc-integrated facular and network magnetic flux,}} $\ftotfn${{,}} by \cite{yeo14a} (black), the reconstruction of this quantity from $\atots$ with the YSK model (red), and the 365-day running mean of the difference between measurement and model (blue). The \cite{balmaceda09} time series goes back to 1874 but the range of the plot is confined to 1974 to 2017, the period where it overlaps with the \cite{yeo14a} time series. The gaps in the \cite{yeo14a} time series correspond to periods where {{there}} no suitable magnetogram observations with which to determine $\ftotfn$ were available. c) {The $H$ term,} which describes the response of $\ftotfn$ to the appearance of a sunspot per unit sunspot area. The dotted line follows the exponential envelope of $H$ and $t_\sun$ denotes the synodic rotation period of the Sun (see Sect. \ref{model1} for details).}
\label{dryskmodel}
\end{figure}

\cite{yeo20} extended the composite time series of daily $\atots$ by \cite{balmaceda09} (Fig. \ref{dryskmodel}a) and of daily $\ftotfn$ by \cite{yeo14a} (Fig. \ref{dryskmodel}b) to June 2017 and constrained the fit parameters with the result. \cite{yeo14a} had derived daily $\ftotfn$ back to 1974 by taking daily full-disc magnetograms from the Kitt Peak Vacuum Telescope \citep{livingston76,jones92}, SoHO/MDI \citep{scherrer95}, and SDO/HMI \citep{scherrer12}, accounting for the systematic differences between the observations from the various telescopes due to instrumental factors and isolating the faculae and { network} present in each magnetogram. YSK could not extend the \cite{balmaceda09} time series beyond June 2017 as it is based on USAF-NOAA sunspot areas for the period of 1986 on, and these measurements are available only up to this time. For $\atots$ in millionths of the solar hemisphere ($\mu$Hem) and $\ftotfn$ in units of Weber (Wb), $h_1=0.843$, $h_2=1.82\times10^{14}{\rm\ Wb}$, $h_3=1.99\times{}10^{11}{\rm\ Wb}$, and $h_4=33.5{\rm\ days}$. The robustness of the model is visibly evident in Fig. \ref{dryskmodel}b; the model reconstruction of $\ftotfn$ from $\atots$ (red) is very similar to the \cite{yeo14a} time series (black), and the 365-day running mean of the difference between the two (blue) is close to null and does not indicate any clear systematic departures. The Pearson's correlation coefficient, $R$, between the two time series is 0.950, meaning the model reproduces about $90\%$ of the observed variability. We refer the reader to \cite{yeo20} for a detailed discussion of the uncertainties in the model and how it compares to earlier models.

The FIR filter $H$ (Equation \ref{eqnysk2}) is based on the earlier findings of PW \citep{preminger05,preminger06a,preminger06b,preminger07}. Sunspots form some time after the BMRs they belong to have started emerging on the solar surface;  they are short-lived   compared to active regions and their decay products, such that the latter persists long after their sunspots have decayed \citep{vandrielgesztelyi15}. Consequently, the response of any quantity that is sensitive to the presence of BMRs, active regions, and their decay products, including $\ftotfn$, to the appearance of a sunspot similarly starts some time before it appears and persists for a period after it dissipates. PW deconvolved $\atots$ from the average unsigned magnetic flux density over the solar disc, solar irradiance, and various chromospheric and coronal indices, deriving FIR filters that elucidate the response of each quantity to the appearance of a sunspot and how {this response} varies with time due to the phenomenon   just described. \cite{preminger07} captured the key features of these empirical FIR filters with a model FIR filter (Equation 2 in their paper), which YSK modified slightly to more accurately reflect the form of the empirical FIR filters, yielding $H$. Depicted in Fig. \ref{dryskmodel}c, $H$ describes a series of sinusoidal lobes, $0.5t_\sun$ wide and $t_\sun$ apart, of diminishing amplitude, as delineated by the exponential envelope of time constant $h_4$ (dotted line). The first lobe has an amplitude of $h_3$ and is centred on $t=0$. The sequence of lobes is truncated at $t=14.25t_\sun$. The FIR filter charts the amount of facular and network magnetic flux associated with a sunspot-bearing BMR, per unit sunspot area, from $0.25t_\sun$ before it fully emerges to $14.25t_\sun$ after. The rising edge of the first lobe corresponds to the BMR emerging, the peak of the first lobe to full emergence, and the falling edge and the subsequent lobes to its decay and modulation with solar rotation.


If the amount of facular and network magnetic flux associated with a sunspot-bearing BMR scales linearly with the area of the enclosed sunspots, then the variation in $\ftotfn$ due to sunspot-bearing BMRs would be given by $\atots\otimes{}H$. However, as noted in the introduction, various studies have found the facular area to increase with increasing sunspot area at a diminishing rate, such that it is at a lower level than if the two quantities are linearly related \citep{foukal93,foukal96,foukal98,chapman97,shapiro14}. This implies that the larger an active region, the higher (lower) the proportion of the total area occupied by sunspots (faculae). Based on this observation, YSK argued that it is unlikely that the amount of facular and network magnetic flux associated with a sunspot-bearing BMR conforms to a linear relationship with the area of the enclosed sunspots. They introduced the fit parameter $h_1$, such that the variation in $\ftotfn$ is given by $\left(\atots\right)^{h_1}\otimes{}H$ instead. This and the modifications introduced to the model FIR filter reported by PW produced a closer agreement between modelled and measured $\ftotfn$, detailed in \cite{yeo20}.

{After the time of the YSK study, the $\atots$ composite by \cite{balmaceda09} was revised by \cite{mandal20}. Both the \cite{balmaceda09} and \cite{mandal20} time series are based on, up to 1976, Royal Greenwich Observatory (RGO) sunspot area measurements. They are therefore exactly identical to one another up to this time, and differ only in the observations used to extend the RGO time series to more recent times. \cite{mandal20} noted that these two time series differ from one another by about 6\% between 1977 and 1985, and that there are otherwise no noticeable systematic differences between the two. The revision of the YSK model with the \cite{mandal20} time series is outside the scope of the current study. In any case, we do not expect the result of applying the \cite{mandal20} time series to the YSK model to differ significantly from what was obtained with the \cite{balmaceda09} time series as input. The disparity between these two time series between 1977 and 1985 is relatively small, especially in comparison to the random error in sunspot area measurements \citep[20\% to 50\%,][]{sofia82}. In addition, less than 10\% of the data points in the daily $\ftotfn$ composite by \cite{yeo14a} lie in this period.}

\subsection{Derivation}
\label{model2}


The total magnetic flux in each newly emerged BMR is the sum of the magnetic flux in its faculae and in its sunspots, or
\begin{equation}
\fbmr=\fbmrf+\fbmrs.
\label{eqnysk3}
\end{equation}
Here we  establish how $\fbmrf$ and $\fbmrs$ relate to $\abmrs$ in turn. To derive the relationship between $\fbmrf$ and $\abmrs$, let us isolate from the YSK model the contribution to $\ftotfn$ by the faculae in newly emerged BMRs, $\ftotbmrf$. As noted in Sect. \ref{model1}, the peak of the first lobe of the FIR filter, which is at $t=0$, corresponds to the response of $\ftotfn$ to a sunspot-bearing BMR when it has fully emerged and has not started to decay. Taking Equation \ref{eqnysk1}, we isolate $\ftotbmrf$ by replacing $H$ with $H\left(t=0\right)=h_3$ and factoring out the basal level, $h_2$, arriving at
\begin{equation}
\ftotbmrf=h_3\left(\atots\right)^{h_1}.
\label{eqnysk4}
\end{equation}
The PW and YSK approach, by modelling the sunspot-related variation in a given disc-integrated quantity as the convolution of $\atots$ with a FIR filter and with the various quantities at daily cadence, treats the appearance of sunspots as one-day events. In other words, it implicitly assumes that each sunspot exists for just one day, such that the total surface coverage by sunspots, $\atots$, is treated in the model as the total surface coverage by sunspots in newly emerged BMRs, $\atotbmrs$. Replacing $\atots$ in Equation \ref{eqnysk4} with $\atotbmrs$, we get
\begin{equation}
\ftotbmrf=h_3\left(\atotbmrs\right)^{h_1}.
\label{eqnysk5}
\end{equation}
Assuming this relationship applies to individual BMRs, then
\begin{eqnarray}
\begin{split}
\fbmrf&=h_3\abmrs^{h_1}\\
&=1.99\times{}10^{11}\abmrs^{0.843}{\rm\ Wb}.
\label{eqnysk6}
\end{split}
\end{eqnarray}
\cite{cho15} examined how the amount of magnetic flux in sunspots and pores varies with their size. The authors noted that for fully formed sunspots, the amount of magnetic flux appears to scale linearly with its area, confirming the early result by \cite{sheeley66}. Assuming the mean magnetic flux density over sunspots is 1350 G or $4.11\times{}10^{11}\ {\rm Wb}\ \mu{\rm Hem^{-1}}$ \citep{solanki93,lites93,keppens96}, this means
\begin{equation}
\fbmrs=4.11\times{}10^{11}\abmrs\ {\rm Wb}.
\label{eqnysk7}
\end{equation}
Taking Equations \ref{eqnysk6} and \ref{eqnysk7} into Equation \ref{eqnysk3}, we get
\begin{eqnarray}
\begin{split}
\fbmr&=4.11\times{}10^{11}\abmrs+1.99\times{}10^{11}{}\abmrs^{0.843}\ {\rm Wb}\\
&=4.11\times{}10^{11}\left(\abmrs+0.485\abmrs^{0.843}\right)\ {\rm Wb}\\
&=4.11\times{}10^{11}\abmrx\ {\rm Wb},
\label{eqnysk8}
\end{split}
\end{eqnarray}
which relates the amount of magnetic flux in newly emerged BMRs to the area of the enclosed sunspots. The $\abmrx$ term, representing $\abmrs+0.485\abmrs^{0.843}$, denotes for a given BMR the area of the sunspot that would contain the same amount of magnetic flux (compare Equations \ref{eqnysk7} and \ref{eqnysk8}). It is worth noting here that the YSK model (Sect. \ref{model1}) and the analysis based on this model that we have just presented pertains only to sunspot-bearing BMRs and does not apply to {ephemeral regions.}

\begin{figure}
\centering
\includegraphics[width=\columnwidth]{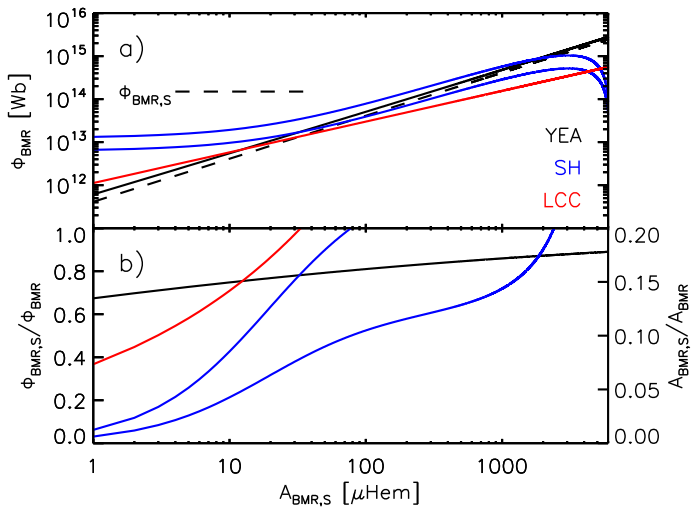}
\caption{{The relationship between BMR magnetic flux and sunspot area.} a) Amount of magnetic flux in newly emerged BMRs, $\fbmr$, as a function of the area of the enclosed sunspots, $\abmrs$, as indicated by the empirical relationship between $\fbmr$ and $\abmrs$ presented in this study (black solid line, Equation \ref{eqnysk8}) and the constraints on $\fbmr$ reported by SH (blue, Equation \ref{eqnsh1}) and LCC (red, Equation \ref{eqnlcc}). For the estimate of $\fbmr$ based on the SH result, the lower and upper bounds, corresponding to $k=1$ and $k=2$, are given. The dashed line follows the amount of magnetic flux in newly emerged BMRs contained in the embedded sunspots, $\fbmrs$ (Equation \ref{eqnysk7}). b) Corresponding plot of $\frac{\fbmrs}{\fbmr}$. To an approximation, $\frac{\fbmrs}{\fbmr}$ is directly proportional to the fraction of the area of each BMR occupied by their sunspots, $\frac{\abmrs}{\abmr}$, indicated on the right axis (Equation \ref{eqnratios}) (see Sects. \ref{model2} and \ref{compareshlcc} for details).}
\label{drratio}
\end{figure}

In Fig. \ref{drratio} we chart $\fbmrs$, $\fbmr$ (dashed and solid black lines, top panel), and the ratio of the two, $\frac{\fbmrs}{\fbmr}$ (black line, bottom panel) as a function of $\abmrs$. As is visibly evident, the empirical relationship between $\fbmr$ and $\abmrs$   presented in Equation \ref{eqnysk8} indicates that most of the magnetic flux in newly emerged BMRs is in the sunspots they enclose. The quantity $\frac{\fbmrs}{\fbmr}$ increases gradually from about 0.7 at $\abmrs=1\ \mu{}{\rm Hem^{-1}}$ to 0.9 at $6000\ \mu{}{\rm Hem^{-1}}$  (left axis, Fig. \ref{drratio}b). \cite{schrijver87} established an empirical relationship between the proportion of the magnetic flux in active regions contained in their sunspots and the proportion of the area of active regions occupied by their sunspots. Assuming it applies to newly emerged BMRs, then
\begin{equation}
\frac{\fbmrs}{\fbmr}=5.1\frac{\abmrs}{\abmr},
\label{eqnratios}
\end{equation}
meaning the trend in $\frac{\fbmrs}{\fbmr}$ with $\abmrs$   (left axis, Fig. \ref{drratio}b) is accompanied by $\frac{\abmrs}{\abmr}$ rising in proportion from about 0.14 to 0.18 (right axis). In other words, the smallest newly emerged BMRs are about 70$\%$ sunspots by magnetic flux and 14$\%$ sunspots by area, and this increases steadily to about 90$\%$ and 18$\%$, respectively for the largest newly emerged BMRs.

We note the following. Firstly, active regions and newly emerged BMRs are  not equivalent to one another, meaning the application of the findings of \cite{schrijver87} to the latter (Equation \ref{eqnratios}) is not strictly correct.  Secondly, some of the smallest BMRs might contain pores instead of  fully developed sunspots \citep[e.g.][]{svanda16}, such that the amount of sunspot magnetic flux they enclose might be smaller than indicated by Equation \ref{eqnysk7}. Nonetheless, the assertion here that the sunspots in a given newly emerged BMR can enclose most of its magnetic flux while occupying only a small proportion of its area is reasonable and within expectation given   that sunspots and pores are magnetically denser than faculae.

\subsection{Comparison to \cite{schrijver94} and \cite{lemerle15}}
\label{compareshlcc}


In this section we compare the empirical relationship between $\fbmr$ and $\abmrs$, derived here (Equation \ref{eqnysk8}) to the constraints on $\fbmr$ reported by SH \citep{schrijver94} and by LCC \citep{lemerle15}. Taking Kitt Peak magnetograms, SH isolated the faculae in {active regions and newly emerged BMRs} and found the facular magnetic flux in individual regions to scale linearly with the area extended by these features. Based on this result and assuming an average BMR magnetic flux density of 100 G or $1.52\times{}10^{10}\ {\rm Wb}\ \mu{\rm Hem^{-1}}$, SH estimated that
\begin{equation}
\fbmr=1.52\times{}10^{10}k\abmr\ {\rm Wb},
\label{eqnsh1}
\end{equation}
where $k$ is a free parameter that the authors did not constrain. SH interpreted the quantity $k-1$ for a given newly emerged BMR as the sum of $\fbmrs$, and the magnetic flux that dissipated before the BMR had fully formed as a proportion of $\fbmr$. Since this ratio is unlikely to exceed unity, at least not by a wide margin, $k$ most probably lies between one and two. Instead of establishing the relationship between $\fbmr$ and $\abmr$ directly, SH had estimated it from the relationship between facular magnetic flux and area. This was necessary because, as noted in precursor studies to the SH study conducted by the same authors \citep{schrijver87,harvey93b}, sunspot magnetic flux cannot be taken directly, at least not reliably, from Kitt Peak magnetograms.

Ideally, to relate $\fbmr$ to $\abmrs$ using the relationship between $\fbmr$ and $\abmr$ reported by SH (Equation \ref{eqnsh1}), it should be combined with the relationship between $\abmr$ and $\abmrs$. Unfortunately, the relationship between $\abmr$ and $\abmrs$ is not known. Instead, let us assume the empirical relationship between the total surface coverage by faculae and by sunspots reported by \cite{chapman97} (hereafter CCD) can be taken as the relationship between the area of the faculae and of the sunspots in individual newly emerged BMRs, such that
\begin{equation}
\abmr=414+22\abmrs-0.0036\abmrs^2.
\label{eqnccd}
\end{equation}
Inserting Equation \ref{eqnccd} into Equation \ref{eqnsh1}, we obtain
\begin{equation}
\fbmr=1.52\times{}10^{10}k\left[414+22\abmrs-0.0036\abmrs^2\right]\ {\rm Wb}.
\label{eqnsh2}
\end{equation}
The estimate of $\fbmr$ from this expression at $k=1$ and $k=2$ represents the lower and upper bound of this quantity based on the combination of the SH and CCD findings.

\cite{lemerle15} compared the $\fbmr$ measurements in the database of newly emerged BMRs compiled by \cite{wang89a} to USAF-NOAA sunspot areas. They could not match each newly emerged BMR in the \cite{wang89a} database to the corresponding sunspot group in the USAF-NOAA record, which excluded a direct comparison and compelled them to compare the distribution of values in the two records instead. From this, LCC estimated that
\begin{equation}
\fbmr=1.12\times{}10^{12}\abmrs^{0.714}\ {\rm Wb}.
\label{eqnlcc}
\end{equation}
Taking Equations \ref{eqnsh2} and \ref{eqnlcc}, we recalculate $\fbmr$ and $\frac{\fbmrs}{\fbmr}$, drawn in Fig. \ref{drratio} (blue and red lines) along the values from the empirical relationship between $\fbmr$ and $\abmrs$ derived in this study (black solid lines).

Figure \ref{drratio}a shows, at lower levels of $\abmrs$, that the estimates of $\fbmr$ based on the SH (blue) and LCC results (red) are higher than the current estimate (black solid line). However, $\fbmr$ increases with increasing $\abmrs$ at a slower rate in the former, such that at higher levels of $\abmrs$ the estimates of $\fbmr$ based on the SH and LCC results drop below  the current estimate and also below $\fbmrs$ (dashed line). This translates into the analogous behaviour of $\frac{\fbmrs}{\fbmr}$ illustrated in Fig. \ref{drratio}b. At low $\abmrs$ the estimates of $\frac{\fbmrs}{\fbmr}$ based on the SH (blue) and LCC results (red) are lower than the current estimate (black), but they rise more rapidly with increasing $\abmrs$ such that they eventually exceed not just the current estimate, but also unity. The amount of magnetic flux in a given BMR has to be greater than the amount carried by the sunspots it encloses, meaning $\fbmr$ going below $\fbmrs$ or, equivalently, $\frac{\fbmrs}{\fbmr}$ going above unity is unphysical. The observation here that the estimates of $\fbmr$ based on the SH and LCC results drop below $\fbmrs$ towards higher $\abmrs$ is unlikely to be from any uncertainty in the latter. The margin is so wide that even if the mean magnetic flux density over sunspots is only one-quarter the value 1350 G we  adopt here, $\fbmrs$ will still be greater than the estimates of $\fbmr$ based on the SH and LCC results for the largest BMRs. The constraint on $\fbmr$ reported by SH, at least when combined with the findings of CCD as we   do here (Equation \ref{eqnccd}), and that by LCC appear to grossly underestimate the amount of magnetic flux in larger BMRs.

The discrepancies between the results reported in this study and earlier by SH and LCC that we have just highlighted could have arisen from the following. SH   examined {active regions and newly emerged BMRs} together even though the two are not equivalent. Furthermore, SH had to constrain $\fbmr$ indirectly as sunspot magnetic flux cannot be determined directly and reliably from the magnetogram observations they   utilised. How the SH results compare to the LCC and current results could have also been affected by the fact that we had to combine the SH result with the findings of CCD in order to make this comparison, and the latter does not strictly apply to newly emerged BMRs.

As for the LCC study, {the authors could not compare the \cite{wang89a} and USAF-NOAA records directly because they could not match each newly emerged BMR in the former to the corresponding sunspot group in the latter.} The \cite{wang89a} database includes spot-free BMRs, which are   not represented in the USAF-NOAA record and should have been excluded. \cite{wang89a} had also determined $\fbmr$   approximately from photographic magnetograms, introducing further uncertainty to the LCC analysis. {SH and LCC were hampered by limitations in the data available to them. As stated in the introduction, it would be straightforward to derive the relationship between $\fbmr$ and $\abmrs$ from an extensive and reliable database of the magnetic and physical properties of newly emerged BMRs, but no such record currently exists. The \cite{wang89a} database is clearly not ideal for this  purpose. The BMR databases compiled more recently by \cite{yeates07} and by \cite{yeates20}, although of excellent quality, are not suitable either as they tabulate the physical properties of each BMR at the time it crosses the central meridian, not when it  newly emerges.}

The YSK model and the analysis based on it presented in this study have their own assumptions and limitations, discussed in Sects. \ref{model1} and \ref{model2}. Nonetheless, it is worth noting that the YSK model allowed us to circumvent the lack of any proper database of newly emerged BMRs, an issue that severely limited the SH and LCC analyses, and constrain{{s}} the relationship between $\fbmr$ and $\abmrs$ from $\ftotfn$ and $\atots$ (i.e. disc-integrated observations) instead.

\section{Application to a surface flux transport model}
\label{sftm}


In this section we  look at the result of applying the empirical relationship between $\fbmr$ and $\abmrs$ presented in this study (Equation \ref{eqnysk8}) and the constraints on $\fbmr$ reported by SH (Equation \ref{eqnsh1}) and by LCC (Equation \ref{eqnlcc}) to the SFTM by CJSS. We generated three surface flux transport simulation runs, denoted  CJSS-SH, CJSS-LCC, and CJSS-YEA (Table \ref{sftmtable}). The CJSS-SH run is generated with the CJSS model as it is and set up following the `reference case' presented by \cite{cameron10}. As we  explain below, the CJSS model (incidentally and not by design) already complies with the constraint on $\fbmr$ reported by SH. The CJSS-LCC and CJSS-YEA runs are generated with the CJSS model modified to comply with the constraints on $\fbmr$ reported by LCC and derived here {{instead}}. In the following, we describe the CJSS model and the simulation set-up  (Sect. \ref{sftm1}) before discussing the results (Sect. \ref{sftm2}).

\begin{table*}
\caption{Surface flux transport simulation runs generated in the current study.}
\label{sftmtable}
\centering
\begin{tabular}{ccccc}
\hline\hline
Simulation run & $\bmaxx\ [{\rm G}\ \mu{}{\rm Hem^{-1}}]$ & $f$ & Constraint on $\fbmr$ & Reference \\
\hline
CJSS-SH  & 0.0156 & $\abmr$         & Equation \ref{eqnsh1}  & \cite{schrijver94} \\
CJSS-LCC & 2.19   & $\abmrs^{0.71}$ & Equation \ref{eqnlcc}  & \cite{lemerle15}   \\
CJSS-YEA & 0.223  & $\abmrx$        & Equation \ref{eqnysk8} & Current study      \\
\hline
\end{tabular}
\tablefoot{The value of $\bmaxx$, the definition of $f$ (Equation \ref{eqncjss8}), and the constraint on $\fbmr$ applied to the run by the definition of $f$ are listed. The total area of each newly emerged BMR, $\abmr$, and the area of the sunspot that would contain the same amount of magnetic flux, $\abmrx$, are related to the area of the enclosed sunspots, $\abmrs$, via Equations \ref{eqnccd} and \ref{eqnysk8}, respectively.}
\end{table*}

\subsection{Simulation set-up}
\label{sftm1}


The governing equation of the CJSS model is
\begin{eqnarray}
\begin{split}
\frac{\partial{}B}{\partial{}t}=&-\Omega\left(\lambda\right)\frac{\partial{}B}{\partial\phi}-\frac{1}{R_{\odot}\cos\lambda}\frac{\partial}{\partial\lambda}\left[\upsilon\left(\lambda\right)B\cos\lambda\right]\\
&+\ns\left[\frac{1}{R^2_{\odot}\cos\lambda}\frac{\partial}{\partial\lambda}\left(\cos\lambda\frac{\partial{}B}{\partial\lambda}\right)+\frac{1}{R^2_{\odot}\cos^{2}\lambda}\frac{\partial^{2}B}{\partial\phi^2}\right]\\
&+D\left(\nr\right)+S\left(\lambda,\phi,t\right),
\label{eqncjss1}
\end{split}
\end{eqnarray}
where $R_\sun$, $\lambda$, and $\phi$ denote the radius of the Sun, latitude, and longitude, respectively. This describes the passive transport of the radial component of the photospheric magnetic field, $B$, under the influence of differential rotation, $\Omega$; meridional flow, $\upsilon$; and diffusion, which is characterised by the surface and radial diffusivity, $\ns$ and $\nr$ \citep{devore84,wang89b,mackay00,schrijver02,baumann04}. Radial diffusion is introduced through a linear operator, $D$ \citep{baumann06}, and BMR emergence through a source term, $S$. As noted in the previous paragraph, we set up the simulation runs following the reference case presented by \cite{cameron10}. We refer the reader to this earlier paper for details on the definition of the initial condition and the $\Omega$, $\upsilon$, $\ns$, and $\nr$ parameters. For the CJSS-SH run, we set up the source term following \cite{cameron10}. For the CJSS-LCC and CJSS-YEA runs, we modified the source term in such a way as to conform the model to the constraints on $\fbmr$ reported by LCC (Equation \ref{eqnlcc}) and derived here (Equation \ref{eqnysk8}). We describe the definition of the source term and these modifications {next}.

\begin{figure}
\centering
\includegraphics{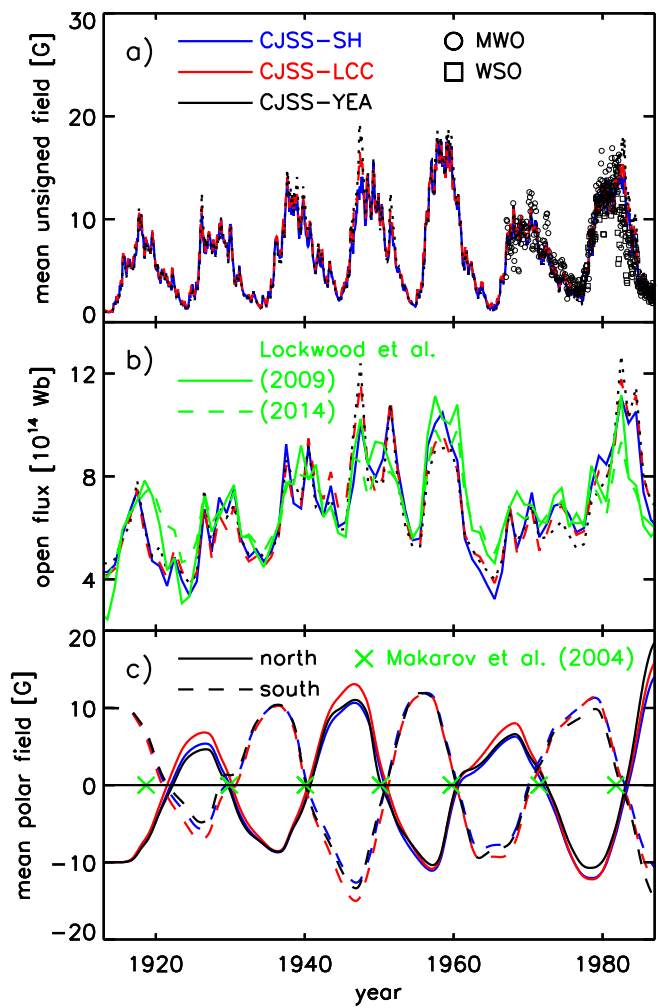}
\caption{{The surface flux transport simulation runs.} a) Average unsigned photospheric magnetic flux density in the CJSS-SH (blue), CJSS-LCC (red), and CJSS-YEA (black) simulation runs (Table \ref{sftmtable}). The CJSS-SH and CJSS-LCC time series are largely hidden by the CJSS-YEA time series due to the close similarity. The circles and squares denote the observed values from MWO and WSO \citep{arge02}. b) Open flux corresponding to each run, derived by extrapolating the model output photospheric magnetic field into the heliosphere, and the reconstruction of this quantity from the $aa$ index by \cite{lockwood09} (green solid line) and by \cite{lockwood14} (green dashed line). {In a) and b) the CJSS-LCC and CJSS-YEA time series are dashed and dotted, respectively, to aid visibility.} c) Mean magnetic field over the north (solid lines) and the south polar caps (dashed lines) in each run. The {green} crosses indicate the observed timing of the polar field reversals reported by \cite{makarov04} (see Sects. \ref{sftm1} and \ref{sftm2} for details).}
\label{drsftm}
\end{figure}

The source term (i.e. details of BMR emergence) is inferred from the composite of the RGO and USAF-NOAA sunspot area records by \cite{hathaway08}. The location and area of each sunspot group when the latter is at its maximum is taken to denote the location and $\abmrs$ of a newly emerged BMR. Following \cite{vanballegooijen98}, each BMR is described as two circular magnetic patches of opposite polarity. Let $B_{+}\latlon$ represent the magnetic flux density of the positive polarity patch at $\latlon$, $\latlonp$ the centre of the patch, and $\beta_{+}\latlon$ the heliocentric angle between $\latlon$ and $\latlonp$, and let $B_{-}\latlon$, $\latlonm$ and $\beta_{-}\latlon$ represent the same for the negative polarity patch. The magnetic flux density profile is given by
\begin{equation}
B_{\pm}\latlon=\bmax\left(\frac{0.4\sep}{\delta}\right)^2\exp\left(-\frac{2\left[1-\cos\beta_{\pm}\left(\lambda,\phi\right)\right]}{\delta^2}\right),
\label{eqncjss4}
\end{equation}
where $\sep=\beta_{+}\latlonm=\beta_{-}\latlonp$ denotes the heliocentric angle between the two patches and $\delta=4^{\circ}$ the size of each patch. The constant factor $\bmax$ modulates the absolute scale of the model output and is fixed at 374 G by matching the output average unsigned magnetic flux density to the observed values from the Mount Wilson Observatory (MWO) and Wilcox Solar Observatory (WSO) \citep{arge02}, illustrated in Fig. \ref{drsftm}a. We note that BMR magnetic flux density, $B_{\pm}$, and therefore the amount of magnetic flux in each BMR, $\fbmr$, are assumed to scale linearly with the square of the angular separation, $\sepsq$. The angular separation, $\sep$, is given by
\begin{equation}
\sep=6.46\times10^{-2}\sqrt{\abmr},
\label{eqncjss5}
\end{equation}
where $\sep$ and $\abmr$ are in units of degrees and $\mu$Hem, respectively, and $\abmr$ is calculated from $\abmrs$ using the CCD relationship (Equation \ref{eqnccd}). \cite{cameron10} established this empirical relationship with the $\sep$ and sunspot umbral area measurements from the Kodaikanal Solar Observatory (KSO) and MWO. The tilt angle of each BMR is calculated from the observed cycle-average tilt angle, also from KSO and MWO, through the procedure detailed in \cite{cameron10}. The source term, and therefore the simulation runs, extend from 1913 to 1986 (solar cycle 15 to 21), the period over which sunspot group location, area, and tilt angle measurements are available. {The simulation runs and the analysis based on them (see  the following sections) are not sensitive to which sunspot area composite we take into the CJSS model as input. The \cite{hathaway08}, \cite{balmaceda09}, and \cite{mandal20} composites are essentially different efforts to extend the RGO record, which ends in 1976, to more recent times with sunspot area measurements from other observatories. The various composites are therefore exactly identical to one another up to this time (i.e. over most of the simulation period of 1913 to 1986).}

The CJSS model relies on the combination of the assumption that $B_{\pm}$ scales linearly with $\sepsq$ (Equation \ref{eqncjss4}) and the empirical relationship between $\sep$ and $\abmr$ (Equation \ref{eqncjss5}) to define $B_{\pm}$. Substituting Equation \ref{eqncjss5} into Equation \ref{eqncjss4} we obtain
\begin{equation}
B_{\pm}\latlon=\bmaxx\abmr\exp\left(-\frac{2\left[1-\cos\beta_{\pm}\left(\lambda,\phi\right)\right]}{\delta^2}\right),
\label{eqncjss7}
\end{equation}
where $\bmaxx=\bmax 0.4^{2}\left(6.46\times10^{-2}\right)^{2}\delta^{-2}=0.0156\ {\rm G}\ \mu{}{\rm Hem^{-1}}$ is a constant factor that serves the same purpose as $\bmax$. The consequence of this approach is that $B_{\pm}$, and therefore also $\fbmr$, scales linearly with $\abmr$ in the model. {Incidentally, this} is similar to the constraint on $\fbmr$ reported by SH, where this quantity also scales linearly with $\abmr$ (Equation \ref{eqnsh1}).

Let us generalise Equation \ref{eqncjss7} by rewriting it as
\begin{equation}
B_{\pm}\latlon=\bmaxx{}f\exp\left(-\frac{2\left[1-\cos\beta_{\pm}\left(\lambda,\phi\right)\right]}{\delta^2}\right),
\label{eqncjss8}
\end{equation}
where $f=\abmr$ determines the $\abmrs$-dependence of $B_{\pm}$ and $\fbmr$. In the CJSS-LCC run we modify the CJSS model to comply with the constraint on $\fbmr$ reported by LCC (Equation \ref{eqnlcc}) by redefining $f$ as $\abmrs^{0.71}$. Similarly, in the CJSS-YEA run we apply the constraint on $\fbmr$ derived in this study (Equation \ref{eqnysk8}) to the CJSS model by redefining $f$ as $\abmrx$. By this redefinition of $f$, we replace the assumption in the CJSS model that $B_{\pm}$ scales linearly with $\sepsq$ with direct observational constraints on the relationship between $\fbmr$ and $\abmrs$. As in the CJSS-SH run, $\bmaxx$ is set in the CJSS-LCC and CJSS-YEA runs at the value that brings the modelled average unsigned magnetic flux density to the absolute scale of the observed values (Fig. \ref{drsftm}a). The value of $\bmaxx$ and the definition of $f$ in each simulation run is summarised in Table \ref{sftmtable}.

\subsection{Simulation results}
\label{sftm2}

\subsubsection{Total photospheric magnetic flux}

{In terms of the total photospheric magnetic flux,} represented in Fig. \ref{drsftm}a by the average unsigned magnetic flux density, the three simulation runs (black, blue, and red curves) are very similar. The agreement in absolute scale is moot since $\bmaxx$ is set in each run at the level that brings the model output to the absolute scale of the observed values (circles and boxes). What is notable is that the time variability is also very similar even with the different $\abmrs$-dependence of $\fbmr$ imposed for each run.

\subsubsection{Solar open flux and polar field reversals}


\cite{cameron10} and \cite{jiang11b} had verified   their model by demonstrating its ability to reproduce the evolution of the solar open flux and the timing of the polar field reversals. The authors derived the open flux corresponding to the model output photospheric magnetic field by extrapolating it into the heliosphere using the Current Sheet Source Surface (CSSS) model by \cite{zhao95a,zhao95b}. This was matched to the independent reconstruction of the open flux from the $aa$ geomagnetic activity index by \cite{lockwood09}. They compared the timing of the polar field reversals in the model output to the timings reported by \cite{makarov04} based on an examination of the polar filaments in H$\alpha$ synoptic charts.


{Following CJSS, we applied the CSSS model to each simulation run to calculate the corresponding solar open flux. In Fig. \ref{drsftm}b we compare the results} (black, blue, and red) to the \cite{lockwood09} reconstruction (green solid line) and the update by \cite{lockwood14} (green dashed line). The various runs do not differ from one another more than they do from the \cite{lockwood09,lockwood14} time series.


In Fig. \ref{drsftm}c we chart the evolution of the polar field in the simulation runs as given by the mean magnetic flux density over the north ($\lambda>75^{\circ}$, solid lines) and the south polar caps ($\lambda<-75^{\circ}$, dashed lines). Polar field reversal is indicated by the mean polar field switching polarity. In the same plot we indicate the observed timing of the polar field reversals reported by \cite{makarov04} ({green} crosses). Analogous to the solar open flux, the various runs do not differ more from one another than they do from the \cite{makarov04} observations. Applying the constraints on $\fbmr$ presented by LCC (Equation \ref{eqnlcc}) and derived in this study (Equation \ref{eqnysk8}) to the CJSS model retained its ability to reproduce the evolution of the solar open flux and the timing of the polar field reversals.

\subsubsection{Correlation between the polar field and solar cycle strength}
\label{babcockleighton}

\begin{figure}
\centering
\includegraphics{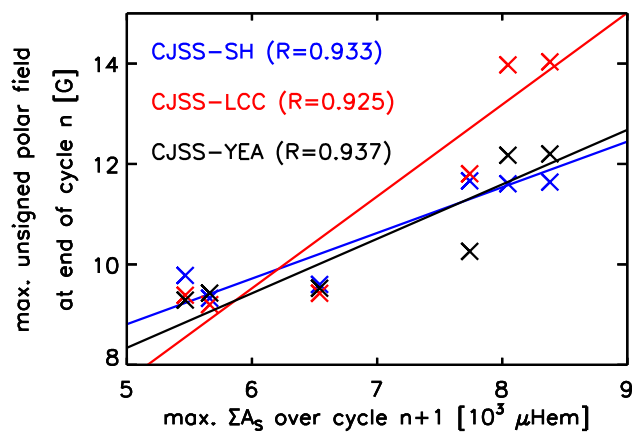}
\caption{For each simulation run, the peak of the mean unsigned magnetic flux density over the polar caps at the end of solar cycle $n$ vs the maximum total surface coverage by sunspots over cycle $n+1$, where $n$ ranges from 15 to 21. We excluded solar cycle 14 as the end of this cycle occurred at the start of the simulation, and is therefore sensitive to the arbitrary initial condition. The crosses and lines denote the individual cycle values and the corresponding linear fit. The Pearson's correlation coefficient, $R$, is indicated (see Sect. \ref{babcockleighton} for details).}
\label{drdynamo}
\end{figure}

A key finding of the CJSS studies is that the amount of polar flux at the end of each solar cycle in the model appears to be highly correlated to the observed strength of the succeeding cycle. This is consistent with the existence of a Babcock--Leighton-type solar dynamo \citep{babcock61,leighton64,leighton69}, where the polar field, which peaks around the end of each cycle, is the main source of the toroidal field in the convection zone \citep[see discussion in][]{cameron15}. We examine if this feature of the CJSS model is sensitive to the $\abmrs$-dependence of $\fbmr$ assumed. Taking each simulation run, we compare the peak of the mean unsigned magnetic flux density over both polar caps at the end of each cycle to the maximum total sunspot area over the following cycle, representing modelled polar flux and measured cycle strength (Fig. \ref{drdynamo}). The Pearson's correlation coefficient, $R$, between the two quantities is 0.933 for the CJSS-SH run, 0.925 for the CJSS-LCC run, and 0.937 for the CJSS-YEA run. The different $\abmrs$-dependence of $\fbmr$ imposed on the various runs made no appreciable difference to the correlation between the model output polar field at the end of each cycle and the observed strength of the succeeding cycle.

Whether in terms of the time variation of the total photospheric magnetic flux (Fig. \ref{drsftm}a) and the solar open flux (Fig. \ref{drsftm}b), the timing of the polar field reversals (Fig. \ref{drsftm}c), or the correlation between the polar field at the end of each cycle and the observed strength of the succeeding cycle (Fig. \ref{drdynamo}), the various simulation runs are very similar. It appears that the CJSS model is, at least to the extent tested here, not particularly sensitive to the $\abmrs$-dependence of $\fbmr$ assumed. This is likely, at least in part, due to the fact that the magnetic flux in each newly emerged BMR can persist for weeks or months, such that the state of a simulation run at a given time depends not only on the BMRs injected then but also on all the BMRs that were introduced in the weeks to months prior. The amount of magnetic flux in individual BMRs depends strongly on the $\abmrs$-dependence of $\fbmr$ assumed. However, the aggregate magnetic flux introduced over an extended period by all the BMRs that had emerged then would not be as sensitive to the $\abmrs$-dependence of $\fbmr$ assumed due to the summation. Nonetheless, it is noteworthy that the application of the empirical relationship between $\fbmr$ and $\abmrs$ presented here to what is an established SFTM retained its key features, including its ability to replicate various independent datasets.

\section{Balance between bright faculae and { network} and dark sunspots at various solar activity levels}
\label{discussion}

As discussed in the Introduction, the balance between bright faculae and { network} and dark sunspots on the solar disc and how that changes with solar activity level is of interest for the relevance to the understanding of the brightness variability of the Sun and other cool stars. Making use of the CJSS-YEA run (Sect. \ref{sftm}), let us examine how the amount of magnetic flux on the solar disc in faculae and { network}, $\ftotfn$, and in sunspots, $\ftots$, relate to one another and how the ratio $\frac{\ftotfn}{\ftots}$ changes with solar activity level. The quantity $\ftots$ is calculated from the \cite{balmaceda09} $\atots$ time series (Fig. \ref{dryskmodel}a) using Equation \ref{eqnysk7}. We obtained the total magnetic flux on the solar disc from the CJSS-YEA run and subtracted
$\ftots$ from it to yield $\ftotfn$. In the following discussion we   examine the annual mean of each quantity.

\begin{figure}
\centering
\includegraphics{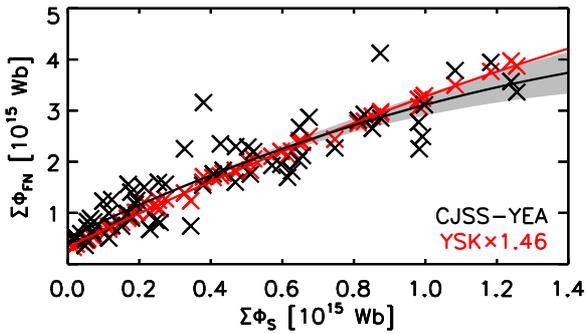}
\caption{The disc-integrated facular and network magnetic flux, $\ftotfn$, as indicated by the CJSS-YEA run (black crosses) and by the YSK model (red crosses), as a function of the disc-integrated sunspot magnetic flux, $\ftots$. The first solar cycle in the CJSS-YEA run (cycle 15, 1913 to 1923) has been excluded as the simulation is sensitive to the arbitrary initial condition here. The annual mean over the period of 1924 to 1986, where there is  data from both models, is shown. The black line and the grey region represents the quadratic polynomial fit to the CJSS-YEA run values and the corresponding 95$\%$ confidence interval, while the red line represents the fit to the YSK model values. See Sect. \ref{discussion} for details.}
\label{dractivity1}
\end{figure}

In Fig. \ref{dractivity1}a we chart $\ftotfn$ estimated from the CJSS-YEA run as a function of $\ftots$ (black crosses). As a check, we   compare this to the YSK model reconstruction of $\ftotfn$ from $\ftots$ (red crosses). Before we discuss the trend in $\ftotfn$ with $\ftots$, let us address the following. Firstly, both the CJSS and YSK models describe the evolution of the photospheric magnetic field due to sunspot-bearing BMRs alone, ignoring ERs. Secondly, the two estimates of $\ftotfn$ differ in absolute scale and  we have to rescale the values from the YSK model by a factor of 1.46 to bring them to the absolute scale of the values from the CJSS-YEA run. Thirdly, plotted against $\ftots$, the values from the CJSS-YEA run are markedly more scattered than the values from the YSK model.

Even though the CJSS model neglects ERs, considering the ability of the CJSS-YEA run to reproduce various independent datasets (Sect. \ref{sftm2}), the effect of this limitation on the corresponding estimate of $\ftotfn$ is likely minor. A similar argument applies to the YSK model estimate of $\ftotfn$. The absolute scale of the CJSS model is set to that of the average unsigned magnetic flux density values reported by \cite{arge02}, which is based on MWO and WSO magnetograms (circles and squares, Fig. \ref{drdynamo}a). The absolute scale of the YSK model is set to that of the $\ftotfn$ time series reported by \cite{yeo14a}, which is based on Kitt Peak, MDI, and HMI magnetograms (black, Fig. \ref{dryskmodel}b). We attribute the different absolute scale of the CJSS and YSK model outputs to systematic uncertainties in how \cite{arge02} and \cite{yeo14a} derived their respective datasets and in the absolute magnetogram signal scale of the various solar telescopes \citep{riley14}. In the YSK model, the variation in $\ftotfn$ due to sunspot-bearing BMRs is given by the convolution of $\atots^{h_1}$ with a FIR filter, $H$ (Equations \ref{eqnysk1} and \ref{eqnysk2}). The implicit assumption is that all BMRs exhibit the same time evolution as delineated by $H$. This ignores the fact that even BMRs containing the same amount of magnetic flux can evolve rather differently due to factors such as latitude, tilt angle, and the magnetic field in their surroundings. These factors are taken into account in the CJSS model, producing the greater scatter in the corresponding $\ftotfn$ versus $\ftots$ plot.

To { more clearly show} the trend in $\ftotfn$ with $\ftots$, we fit a quadratic polynomial in $\ftots$ to each of the two estimates of $\ftotfn$. The fit to the CJSS-YEA run values (black line, Fig. \ref{dractivity1}) is given by
\begin{equation}
\begin{aligned}
\ftotfn=&\left(0.52\pm0.11\right)+\left(3.33\pm0.53\right)\ftots\\
&-\left(0.73\pm0.46\right)\ftots^2
\end{aligned}
\end{equation}
and the fit to the YSK model values (red line) by
\begin{equation}
\begin{aligned}
\ftotfn=&\left(0.334\pm0.019\right)+\left(3.367\pm0.086\right)\ftots\\
&-\left(0.425\pm0.074\right)\ftots^2.
\end{aligned}
\end{equation}
This revealed that $\ftotfn$  increases with increasing $\ftots$ at a diminishing rate such that it appears to gradually saturate, analogous to what earlier studies noted comparing facular area {and various disc-integrated quantities that are sensitive to the amount of faculae and { network} present} to sunspot area and number \citep{foukal93,foukal96,foukal98,chapman97,solanki98,shapiro14,yeo20}, discussed in the Introduction. The fit to the CJSS-YEA run values (black line) and the fit to the YSK model values (red line) do differ slightly, though the latter  lies within the 95$\%$ confidence interval of the former (shaded region). This means {that}, taking the scatter in the CJSS-YEA run values (black crosses) into account, {{that}} there is little separating the two estimates of $\ftotfn$ in terms of the $\ftots$-dependence. The CJSS-YEA run and the YSK model are not entirely independent of one another as we had constrained the former using the empirical relationship between $\fbmr$ and $\abmrs$ derived from the latter. Nonetheless, it is significant that two models that adopt vastly different approaches to modelling the evolution of the photospheric magnetic field returned similar results.

\begin{figure*}
\centering
\includegraphics{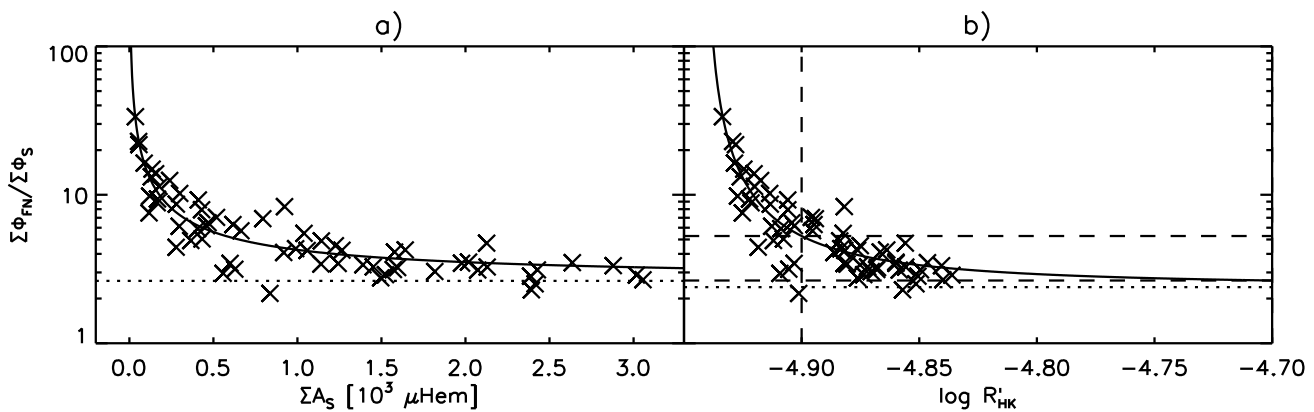}
\caption{{The variation in the balance between bright faculae and network and dark sunspots with solar activity.} a) The ratio $\frac{\ftotfn}{\ftots}$, where $\ftotfn$ is obtained from the CJSS-YEA run, as a function of $\atots$ (crosses). As in Fig. \ref{dractivity1}, the annual mean over the period of 1924 to 1986 is depicted. The solid line follows the power-law fit (Equation \ref{eqnact2}) and the dotted line gives the position of the horizontal asymptote as given by the constant term in this fit. b) The same, but for $\frac{\ftotfn}{\ftots}$ as a function of $\logrhk$. The vertical dashed line indicates $\logrhk=-4.9$ and the horizontal dashed lines gives the value of $\frac{\ftotfn}{\ftots}$ at $\logrhk=-4.9$ and $\logrhk=-4.7$, as given by the power-law fit (Equation \ref{eqnact3}) (see Sect. \ref{discussion} for details).}
\label{dractivity2}
\end{figure*}

Next, we examine what the apparent trend in $\ftotfn$ with $\ftots$ implies for the balance between the amount of faculae and { network} and the amount of sunspots on the solar disc and how that changes with solar activity level. Sunspot indices are often employed as proxies of solar activity, while stellar activity is often characterised by the quantity $\logrhk$ defined in \cite{noyes84}. Taking this into consideration, let us examine $\frac{\ftotfn}{\ftots}$, as estimated from the CJSS-YEA run, as a function of $\atots$ and of $\logrhk$ (crosses, Fig. \ref{dractivity2}). We make use of the {model reconstruction of $\logrhk$ from $\atots$ presented by YSK.}

The gradual saturation in $\ftotfn$ as we go from low to high $\ftots$ (Fig. \ref{dractivity1}) translates into $\frac{\ftotfn}{\ftots}$ decreasing with increasing solar activity level at a diminishing rate such that it appears to approach a horizontal asymptote (Fig. \ref{dractivity2}). Let us model $\frac{\ftotfn}{\ftots}$ as conforming to a power-law relationship with activity level such that
\begin{equation}
\frac{\ftotfn}{\ftots}=g_1\left[\alpha+g_2\right]^{g_3}+g_4,
\label{eqnact1}
\end{equation}
where $\alpha$ represents activity as given by either $\atots$ or $\logrhk$ and $g_1$ to $g_4$ are fit parameters. The $g_1\left[\alpha+g_2\right]^{g_3}$ term describes the variation in $\frac{\ftotfn}{\ftots}$ with $\alpha$, while $g_4$ denotes the position of the horizontal asymptote. We fit Equation \ref{eqnact1} to the $\frac{\ftotfn}{\ftots}$ versus $\atots$ plot (Fig. \ref{dractivity2}a) and the $\frac{\ftotfn}{\ftots}$ versus $\logrhk$ plot (Fig. \ref{dractivity2}b) in turn. For the former, $g_2$ is fixed at null. At convergence we arrive at
\begin{equation}
\frac{\ftotfn}{\ftots}=\left(690\pm110\right)\atots^{\left(-0.876\pm0.043\right)}+\left(2.63\pm0.29\right)
\label{eqnact2}
\end{equation}
and
\begin{equation}
\begin{aligned}
\frac{\ftotfn}{\ftots}&=\\
&\left(0.038\pm0.060\right)\left[\logrhk+\left(4.9405\pm0.0046\right)\right]^{\left(-1.35\pm0.45\right)}\\
&+\left(2.39\pm0.72\right),
\end{aligned}
\label{eqnact3}
\end{equation}
represented in Fig. \ref{dractivity2} by the solid lines. In both instances the $g_4$ term or the position of the horizontal asymptote (dotted lines) is above unity, implying that the amount of magnetic flux on the solar disc in sunspots never exceeds that in faculae and { network}. {This is} consistent with \cite{parnell09} and \cite{thornton11}, who came to a similar conclusion {by examining the distribution of magnetic features} with the amount of flux in each feature.

The observation here that most of the magnetic flux on the solar disc is in faculae and { network}  does not contradict our earlier assertion that most of the magnetic flux in newly emerged BMRs is in the enclosed sunspots (Sect. \ref{model2}). The magnetic flux in BMRs persists long after the comparatively short-lived sunspots they enclose have dissipated, and {manifests} as faculae and { network} \citep{vandrielgesztelyi15}. {In combination with the fact that} the photospheric magnetic field at a given time is comprised of not just the BMRs that had emerged then, but also what remains from all the BMRs that had emerged in the weeks to months prior, the two assertions we are making do not exclude one another.

In the context of the brightness variability of cool stars, the transition between the faculae-dominated and spot-dominated regimes is estimated to be in the range of $\logrhk$ of $-4.9$ to $-4.7$ \citep{lockwood07,hall09,shapiro14,radick18,reinhold19}. Looking at Fig. \ref{dractivity2}b, the $\logrhk$ of the Sun (crosses) is in the range of $-4.93$ to $-4.84$, around the lower end of the range of activity where this transition occurs. The extrapolation of the power-law fit (solid line) indicates that in the range of $\logrhk$ of $-4.9$ to $-4.7$, $\frac{\ftotfn}{\ftots}$ varies slowly   with activity, such that it stays within a relatively narrow range of about 3 to 5 (dashed lines). Taken together, these two observations imply that the $\frac{\ftotfn}{\ftots}$ level of the Sun might not be far above the level where facular and network brightening and sunspot darkening would cancel one another out. This concurs with the findings of  \cite{shapiro16} and \cite{radick18}, who noted that while the Sun is faculae-dominated, it is only marginally so as facular and network brightening and sunspot darkening appear to be closely balanced.

\section{Summary}
\label{summary}

The relationship between sunspot-bearing bipolar magnetic regions (BMRs) and the sunspots they enclose, connected to how solar magnetism and activity relate to sunspot prevalence, is important. {It is particularly relevant to surface flux transport models (SFTMs) that derive the details of BMR emergence they require as input from sunspot observations.} However, attempts to constrain how the amount of magnetic flux in each newly emerged BMR, {$\fbmr$}, relates to its area \citep{schrijver94} or the area of the enclosed sunspots, $\abmrs$ \citep{lemerle15}, were limited by the fact that there is no extensive and reliable database of the magnetic and physical properties of newly emerged BMRs currently available. In the absence of proper observational constraints, existing SFTMs adopt various simplifications to determine the details of BMR emergence from sunspot observations. In this study we aimed to establish the relationship between $\fbmr$ and $\abmrs$.

{To this end,} we made use of the empirical model of the relationship between the disc-integrated facular and network magnetic flux, $\ftotfn$, and the total surface coverage by sunspots, $\atots$, presented by \cite{yeo20}. {The structure of the \cite{yeo20} model enabled us to circumvent the lack of any proper BMR database, and instead derive an empirical relationship between $\fbmr$ and $\abmrs$ (Equation \ref{eqnysk8}) from $\atotfn$ and $\atots$ (i.e. disc-integrated quantities).} The empirical relationship between $\fbmr$ and $\abmrs$ indicates that for newly emerged BMRs, the enclosed sunspots occupy a small proportion of their area (14$\%$ to 18$\%$), but contain most of their magnetic flux (70$\%$ to 90$\%$) (Fig. \ref{drratio}b). We found evidence that the {earlier results} by \cite{schrijver94} (Equation \ref{eqnsh1}) and \cite{lemerle15} (Equation \ref{eqnlcc}) might underestimate the $\fbmr$ of larger BMRs (Fig. \ref{drratio}a).

To verify the empirical relationship between $\fbmr$ and $\abmrs$ presented here, we examined the result of applying this and the earlier results by \cite{schrijver94} and \cite{lemerle15} to a SFTM. We made use of the SFTM reported by \cite{cameron10} and \cite{jiang11a,jiang11b}. This model, incidentally and not by design, already complies with the \cite{schrijver94} {result.} We generated three simulation runs, one with the model as it is and the other two with the model modified to comply with the {\cite{lemerle15} results and the current results}. We compared the evolution of the solar open flux to the independent reconstruction of this quantity from the $aa$ geomagnetic activity index by \cite{lockwood09,lockwood14} (Fig. \ref{drsftm}b) and the timing of the polar field reversals to the observed timing reported by \cite{makarov04} (Fig. \ref{drsftm}c). The various runs {reproduce the \cite{lockwood09,lockwood14} and \cite{makarov04} results relatively well and almost equally.} An important feature of the \cite{cameron10} and \cite{jiang11a,jiang11b} model is that the polar field at the end of each cycle is highly correlated to the observed strength of the following cycle. This is significant as it is consistent with the existence of a Babcock--Leighton-type solar dynamo. In terms of this correlation, the various runs are again very similar (Fig. \ref{drdynamo}). Surface flux transport simulations do not appear to be, to the extent tested here, particularly sensitive to the $\abmrs$-dependence of $\fbmr$ assumed. {Nonetheless, it is worth noting that applying the current result to what is an established SFTM retained its key features.}

Making use of the surface flux transport simulation based on the relationship between $\fbmr$ and $\abmrs$ presented here, we examined how the amount of magnetic flux on the solar disc in faculae and { network}, $\ftotfn$, and in sunspots, $\ftots$, relate to one another and the relationship between the ratio $\frac{\ftotfn}{\ftots}$ and the solar activity level. We found $\ftotfn$ to increase with increasing $\ftots$ at a diminishing rate such that it appears to gradually saturate (Fig. \ref{dractivity1}). {This complies with the analogous results from the earlier studies that compared facular area and various disc-integrated quantities sensitive to the amount of faculae and { network} present to sunspot area and number} \citep{foukal93,foukal96,foukal98,chapman97,solanki98,shapiro14,yeo20}. Taking $\atots$ and $\logrhk$ as proxies of solar activity, we found $\frac{\ftotfn}{\ftots}$ to decrease with increasing activity at a diminishing rate such that it appears to approach a horizontal asymptote (Fig. \ref{dractivity2}). We modelled $\frac{\ftotfn}{\ftots}$ as conforming to a power-law relationship with activity (Equations \ref{eqnact2} and \ref{eqnact3}). The power-law model indicates that, at least within the range of activity examined, the proportion of the magnetic flux on the solar disc that is in sunspots remains smaller than that in faculae and { network}. This does not contradict our claim that most of the magnetic flux in newly emerged BMRs is in the enclosed sunspots (see  Sect. \ref{discussion}). The extrapolation of the power-law model to higher activity levels indicates that the $\frac{\ftotfn}{\ftots}$ level of the Sun might be close to the level where facular and network brightening and sunspot darkening would cancel one another out, {consistent with \cite{shapiro16} and \cite{radick18}.}

\begin{acknowledgements}
The authors are grateful to R. H. Cameron of the Max-Planck Institut f\"{u}r Sonnensystemforschung for the useful discussions and for granting access to the surface flux transport model featured in this work. KLY and NAK received funding from the German Federal Ministry of Education and Research (project 01LG1909C), KLY and SKS from the European Research Council through the Horizon 2020 research and innovation program of the European Union (grant 695075) and JJ from the National Science Foundation of China (grant 11873023 and 11522325) and the Fundamental Research Funds for the Central Universities of China.
\end{acknowledgements}

\bibliographystyle{aa}
\bibliography{references}

\end{document}